\documentclass[reprint,twocolumn,aps,floatfix,prb,showpacs]{revtex4-1}

\usepackage{amsmath}
\usepackage{amssymb}
\usepackage{graphicx}
\usepackage{dcolumn}
\usepackage{bm}
\usepackage[colorlinks=true, allcolors=blue]{hyperref}
\usepackage{epstopdf}
\usepackage[utf8]{inputenc}
\usepackage{amssymb}
\usepackage{hyperref}

\begin{document}

\title{Doping dependence of superconductivity on a honeycomb lattice within the framework of kinetic-energy-driven superconductivity}

\author{Yu Lan}
\email{Correspondence: ylan@bnu.edu.cn}

\author{Xian-Feng Yu}

\author{Li-Ting Zhang}

\affiliation{College of Physics and Electronic Engineering, Hengyang Normal University, Hengyang 421002, China}

\begin{abstract}
Unconventional superconductivity on a honeycomb lattice has received increasing interest since the discovery of graphene primarily due to the similarities between materials with a honeycomb lattice and cuprate superconductors. Many theoretical studies have been conducted on superconductivity on a honeycomb lattice, however, a consistent picture is still lacking. In this article we have extended the theory of kinetic-energy-driven superconductivity, which has been developed to investigate unconventional superconductivity in cuprate superconductors, to explore superconductivity on a honeycomb lattice within the $t$-$J$ model. Our results demonstrate that the charge-carrier pair gap parameter with $d_{x^{2}-y^{2}}+{\rm i}d_{xy}$-wave symmetry exhibits a dome-like shape as a function of doping, with superconductivity emerging at a certain doping concentration and disappearing at high doping levels, similar to what has been observed in cuprate and cobaltate superconductors. Furthermore, the charge-carrier pair gap parameter decreases with increasing the value of $J/t$ (the antiferromagnetic exchange coupling constant relative to the nearest-neighbor hopping integral), and approaches zero when $J/t$ reaches a sufficiently large value. This indicates that the antiferromagnetic order will suppress the superconducting state and a sufficiently strong exchange coupling will completely destroy the superconductivity. Taking into account our present results together with the corresponding results of cuprate and cobaltate superconductors, it appears that the dome-like shape of the doping dependence of the charge-carrier pair gap parameter may be a common feature in doped Mott insulators.
\end{abstract}

\maketitle


Unconventional superconductivity has attracted great interest since the discovery of cuprate superconductors \cite{Bednorz86}. Cuprate superconductors are quasi-two-dimensional materials with the main physics occurring in the CuO$_{2}$ planes, which have a square lattice structure. The parent compounds of cuprate superconductors are half-filling antiferromagnetic (AF) Mott insulators. When the CuO$_{2}$ planes are doped away from half filling, the AF long-range order (AFLRO) is suppressed leaving the AF short-range correlation still intact \cite{Kastner98,Fujita12}. Then superconductivity emerges at a doping concentration that is not too far from half filling. One noteworthy characteristic of cuprate superconductors is that both the superconducting (SC) gap parameter $\Delta$ and the critical temperature $T_{\rm c}$ exhibit a dome-like shape with respect to doping concentration \cite{Kordyuk10}. In the SC doping range, electron Cooper pairs form and condensate below $T_{\rm c}$. Unlike conventional superconductors, cuprate superconductors belong to strongly correlated electron systems (SCES), and Cooper pairs are not bound together by electron-phonon interactions. As a result, the understanding of the SC mechanism in cuprate superconductors remains quite complex and controversial till now.

The successful preparation of graphene \cite{Geim04} offers another platform to study the physics of SCES and unconventional superconductivity. Graphene is the first discovered real two-dimensional (2D) material with the C atoms being arranged into a honeycomb lattice. Materials with the honeycomb lattice have some similarities with cupate superconductors, thus gaining increasing attention on exploration their superconductivity. Firstly, most materials with the honeycomb lattice and cuprate superconductors with the square lattice both belong to 2D SCES. Secondly, the honeycomb lattice theoretically supports an AF state at half filling in the strong-coupling limit, which is similar to the undoped CuO$_{2}$ square lattice. According to the results from the Hubbard model \cite{Assaad13,Otsuka16}, the half-filled honeycomb lattice changes from a semimetal to a Mott insulator for strong interaction with sufficiently large on-site Coulomb repulsion $U$. And it was revealed by experiments that the honeycomb lattice material In$_{3}$Cu$_{2}$VO$_{9}$ is an AF insulator at half filling \cite{Kriener08,Chen12} and considered to be SCES. Thirdly, many theoretical studies demonstrated that the AFLRO on the honeycomb lattice is suppressed \cite{Paiva05,Herbut06,Raghu08,Meng10,Gu13,Assaad13,Arya15,Zhong15,Plakida18,Khee-Kyun20,Ribeiro23} and the SC state rises \cite{Uchoa07,Black-Schaffer07,Honerkamp08,Jiang08,Pathak10,Raghu10,Pellegrino10,Ma11,Nandkishore12nphy,Nandkishore12prb,Wang12,Kiesel12,Wu13,Gu13,Black-Schaffer14,Jiang14,Ma14,Faye15,Xu16,Li16,Ying18,Plakida19,Guo19,Qi20,Gu20,Song-Jin21,Ying20,Ying22,Qin22,Yu22,Ribeiro23,Khee-Kyun20,Khee-Kyun23} upon doping charge carriers into the lattice, which is similar to the results of cuprate superconductors. Experimentally, though till now no real observation of superconductivity in pure monolayer graphene has been reported, phonon-driven conventional superconductivity has already been observed in the intercalated graphites and similar materials. In early days after the discovery of graphene, alkali metal–graphite intercalation compounds were found to exhibit superconductivity \cite{Weller05,Csanyi05}. Later, the presence of superconductivity was reported in K-doped few-layer graphene \cite{Xue12}, Ca-intercalated bilayer graphene \cite{Ichinokura16}, Li-decorated monolayer graphene \cite{Ludbrook15} and Ca-doped graphene laminates \cite{Chapman16}. And it was suggested that graphene contacted with SC electrodes can induce superconductivity through the proximity effect \cite{Heersche07,Du08,Ojeda-Aristizabal09,Lee18}. Although superconductivity in these graphene-related materials is primarily induced by electron-phonon interaction, their SC properties are believed to be strongly influenced by unconventional non-adiabatic effects as demonstrated in the representative case of Li-decorated graphene \cite{Szcz19,Szcz23}, which shares similarities with the results of bismuthates \cite{Szcz21}. It is noteworthy that, although $T_{\rm c}$ of these materials is relatively low, the upper theoretical limit for the honeycomb-type lattice is estimated to be around 36 K \cite{Drzazga18}. One exciting discovery is that unconventional superconductivity was recently achieved in doped magic-angle-twisted bilayer graphene \cite{Cao18sc} while the compound at half filling is a Mott insulator due to the very flat electronic band near the Fermi energy \cite{Cao18}. Other materials with honeycomb lattice, like pnictide SrPtAs \cite{Nishikubo11,Biswas13} and $\beta$-MNCl (M = Hf, Zr) \cite{Yamanaka96,Yamanaka98}, have also been observed to exhibit unconventional superconductivity.

For superconductivity the paring symmetry is one of the most important properties. Being similar to the CuO$_{2}$ square lattice as mentioned above, the $d$-wave symmetry superconductivity on the honeycomb lattice is naturally considered. However, the honeycomb lattice belongs to the hexagonal $D_{6h}$ group with $k_{z}=0$, while the fourfold-symmetric $d$-wave solutions belong to a 2D irreducible representation $E_{2g}$, thus leading to the necessity of two $d$-wave solutions or the linear combination of them to eliminate the incompatibility. As a natural choice, the chiral $d_{x^{2}-y^{2}}+{\rm i}d_{xy}$-wave (thereafter simplified as $d+{\rm i}d$) symmetry is generally expected. Many theories have claimed that $d+{\rm i}d$-wave symmetry dominates the superconductivity in a certain doping range, such as functional renormalization group (FRG) \cite{Honerkamp08,Raghu10,Kiesel12,Wu13}, truncated unity FRG \cite{Song-Jin21}, random phase approximation (RPA) \cite{Li16}, variational Monte Carlo calculations \cite{Pathak10,Wang12,Jiang14,Yu22}, quantum Monte Carlo (QMC) simulations \cite{Ma11,Black-Schaffer14,Ying18,Guo19,Ying20,Ying22}, density matrix renormalization group \cite{Jiang14}, renormalized mean-field theory \cite{Black-Schaffer14}, Grassmann tensor product state (GTPS) approach \cite{Gu13}, and other theories \cite{Jiang08,Nandkishore12nphy,Xu16,Plakida19,Khee-Kyun20,Ribeiro23,Khee-Kyun23}. Other pairing symmetries have also been proposed. For example, the $p_{x}\pm{\rm i}p_{y}$-wave pairing state has been claimed by FRG \cite{Raghu10}, QMC calculations \cite{Ma14,Ying20}, variational cluster approximation \cite{Faye15}, large-scale dynamic cluster approximation \cite{Xu16}, and GTPS approach \cite{Gu20}, etc. The tripled intra-sublattice $f$-wave pairing was also proposed by RPA \cite{Li16}, FRG \cite{Kiesel12}, renormalization group scheme \cite{Honerkamp08}, and so on.

Although the theoretical investigations on superconductivity on the honeycomb lattice are rather fruitful, due to the lack supporting of experimental results, the SC properties especially the pairing symmetries are still under intense debate and far from reaching a consistent picture. Furthermore, most of the results mainly originate from numerical calculations with few microscopic theories, so it is still necessary to further study the superconductivity of materials with the honeycomb lattice to understand their SC properties and SC mechanism. On the other hand, studying the doped Mott insulators on a honeycomb lattice itself can provide a rich playground for novel physical properties of unconventional superconductors. Hence, in this article we try to extend the theory of kinetic-energy-driven superconductivity \cite{Feng03,Feng15}, which has been developed to study the SC properties of cuprate superconductors, to explore the unconventional superconductivity on a honeycomb lattice.

It is widely believed that the $t$-$J$ model \cite{Zhang88} is able to capture the basic physics of doped Mott insulators \cite{Anderson87,Zhang88,Phillips10,Feng9404,Feng04,Feng03,Feng15}. The similarities between materials with a honeycomb lattice and cuprate superconductors with a square lattice have led to the adoption of the $t$-$J$ model for studying doped honeycomb lattices \cite{Wu13,Gu13,Black-Schaffer14,Zhong15,Feng16,Plakida18,Plakida19,Khee-Kyun20,Khee-Kyun23}. As the honeycomb lattice is a bipartite lattice, the $t$-$J$ model for this lattice can be written as \cite{Feng16},
\begin{eqnarray}\label{tjham}
H&=&-t\sum_{l\hat{\eta}\sigma}(C^{\dagger}_{{\rm A}l\sigma}C_{{\rm B}l+\hat{\eta}\sigma}+C^{\dagger}_{{\rm B}l\sigma}C_{{\rm A}l+\hat{\eta}\sigma})+\mu\sum_{\upsilon l\sigma} C^{\dagger}_{\upsilon l\sigma}C_{\upsilon l\sigma} \nonumber\\
&&+J\sum_{l\hat{\eta}}({\bf S}_{{\rm A}l}\cdot {\bf S}_{{\rm B}l+\hat{\eta}}+{\bf S}_{{\rm B}l}\cdot {\bf S}_{{\rm A}l+\hat{\eta}}), ~~~~~~~
\end{eqnarray}
supplemented by a local constraint $\sum_{\sigma}C^{\dagger}_{\upsilon l\sigma} C_{\upsilon l\sigma}\leq 1$ to avoid double electron occupancy. The subscripts A and B denote two different sublattices, $l+\hat{\eta}$ represent the nearest-neighbor (NN) sites of $l$, $t$ is the NN hopping integral, $J$ is the NN spin-spin AF exchange coupling constant, and $\mu$ is the chemical potential. $C^{\dagger}_{\upsilon l\sigma}$ and $C_{\upsilon l\sigma}$ ($\upsilon= {\rm A,B}$ and $\sigma=\uparrow, \downarrow$) are electron creation and annihilation operators, respectively, while ${\bf S}_{\upsilon l}=(S^{\rm x}_{\upsilon l},S^{\rm y}_{\upsilon l}, S^{\rm z}_{\upsilon l})$ are spin operators. For cuprate superconductors, the local constraint of no double electron occupancy can be treated properly within the charge-spin separation (CSS) fermion-spin theory \cite{Feng9404,Feng04,Feng15}. Following this theory, the constrained electron operators $C_{\upsilon l\uparrow}$ and $C_{\upsilon l\downarrow}$ can be decoupled as,
\begin{eqnarray}\label{fermion-spin-theory}
C_{\upsilon l\uparrow}=h^{\dagger}_{\upsilon l\uparrow}S^{-}_{\upsilon l},~~~~C_{\upsilon l\downarrow}=h^{\dagger}_{\upsilon l\downarrow}S^{+}_{\upsilon l},
\end{eqnarray}
respectively, where the spin operators $S^{\pm}_{\upsilon l}=S^{\rm x}_{\upsilon l}\pm{\rm i}S^{\rm y}_{\upsilon l}$ represent the spin degree of freedom, while the fermion operator $h_{\upsilon l\sigma}=e^{-i\Phi_{l\sigma}}h_{\upsilon l}$ keeps track of the charge degree of freedom together with some effects of spin configuration rearrangements due to the presence of the doped charge carrier itself, and then the local constraint of no double electron occupancy is always satisfied in actual calculations. Within this representation (\ref{fermion-spin-theory}), the original $t$-$J$ model (\ref{tjham}) can be rewritten as,
\begin{eqnarray}\label{cssham}
H&=&t\sum_{l\hat{\eta}}(h^{\dagger}_{{\rm B}l+\hat{\eta}\uparrow}h_{{\rm A}l\uparrow}S^{+}_{{\rm A}l}S^{-}_{{\rm B}l+\hat{\eta}}+h^{\dagger}_{{\rm B}l+\hat{\eta}\downarrow}
h_{{\rm A} l\downarrow}S^{-}_{{\rm A}l} S^{+}_{{\rm B}l+\hat{\eta}}\nonumber\\
&&+h^{\dagger}_{{\rm A}l+\hat{\eta}\uparrow}h_{{\rm B}l\uparrow}S^{+}_{{\rm B}l}S^{-}_{{\rm A}l+\hat{\eta}} +h^{\dagger}_{{\rm A}l+\hat{\eta}\downarrow}h_{{\rm B}l\downarrow} S^{-}_{{\rm B}l} S^{+}_{{\rm A}l+\hat{\eta}})\nonumber\\
&&-\mu\sum_{\upsilon l\sigma}h^{\dagger}_{\upsilon l\sigma}h_{\upsilon l\sigma}\nonumber\\
&&+J_{{\rm eff}}\sum_{l\hat{\eta}}({\bf S}_{{\rm A}l}\cdot{\bf S}_{{\rm B}l+\hat{\eta}}
+{\bf S}_{{\rm B}l}\cdot{\bf S}_{{\rm A}l+\hat{\eta}}),~~~~~
\end{eqnarray}
where $J_{{\rm eff}}=(1-\delta)^{2}J$, and
$\delta=\langle h^{\dagger}_{\upsilon l\sigma}h_{\upsilon l\sigma}\rangle=\langle h^{\dagger}_{\upsilon l}h_{\upsilon l}\rangle$ is the doping concentration. Within the CSS $t$-$J$ Hamiltonian (\ref{cssham}), the mean-field (MF) charge-carrier and spin Green's functions have been obtained. The intra-sublattice and inter-sublattice parts of the MF charge-carrier Green's functions are,
\begin{subequations}\label{MF-charge-Green-function}
\begin{eqnarray}
g^{(0)}_{\rm intra}({\bf k},\omega)&=&{1\over 2}\sum_{\nu=1,2}{1\over \omega-\xi^{(\nu)}_{\bf k}}, \label{MF-charge-intra-part}\\
g^{(0)}_{\rm inter}({\bf k},\omega)&=&{1\over 2}e^{i\theta_{\bf k}}\sum_{\nu=1,2}(-1)^{\nu+1}{1\over \omega-\xi^{(\nu)}_{\bf k}},\label{MF-charge-inter-part}
\end{eqnarray}
\end{subequations}
where the MF charge-carrier excitation spectrum $\xi^{(\nu)}_{\bf k}=(-1)^{\nu+1}Zt\chi|\gamma_{\bf k}|-\mu$, the spin correlation function $\chi=\langle S^{+}_{{\rm A}l}S^{-}_{{\rm B}l+\hat{\eta}}\rangle=\langle S^{+}_{{\rm B} l}S^{-}_{{\rm A}l+\hat{\eta}}\rangle$, $Z=3$ is the number of the NN sites on a honeycomb lattice, $\gamma_{\bf k}=(1/Z)\sum_{\hat{\eta}} e^{-i{\bf k}\cdot {\hat{\eta}}}=e^{i\theta_{\bf k}}|\gamma_{\bf k}|$. The intra-sublattice and inter-sublattice parts of the MF spin Green's functions are,
\begin{subequations}\label{MF-spin-Green-function}
\begin{eqnarray}
D^{(0)}_{\rm intra}({\bf k},\omega)&=&\sum_{\nu=1,2}{B^{(\nu)}_{\bf k}\over 2\omega^{(\nu)}_{\bf k}}\left ({1\over \omega-\omega^{(\nu)}_{\bf k}}-{1\over\omega+\omega^{(\nu)}_{\bf k}} \right ), ~~~~~~ \label{MF-spin-GF-intra-part-1}\\
D^{(0)}_{\rm inter}({\bf k},\omega)&=&e^{i\theta_{\bf k}}\sum_{\nu=1,2}(-1)^{\nu+1}{B^{(\nu)}_{\bf k}\over 2\omega^{(\nu)}_{\bf k}}\nonumber\\
&&\times\left ( {1\over \omega-\omega^{(\nu)}_{\bf k}} - {1\over \omega+\omega^{(\nu)}_{\bf k}}\right ), \label{MF-spin-GF-inter-part-1} \\
D^{(0)}_{\rm (z)intra}({\bf k},\omega)&=&\sum_{\nu=1,2}{B^{(\nu)}_{{\rm z}{\bf k}}\over 2\omega^{(\nu)}_{{\rm z}{\bf k}}}\left ({1\over \omega-\omega^{(\nu)}_{{\rm z}{\bf k}}}- {1\over \omega+\omega^{(\nu)}_{{\rm z}{\bf k}}}\right ),\label{MF-spin-GF-intra-part-2}\\
D^{(0)}_{\rm (z)inter}({\bf k},\omega)&=&e^{i\theta_{\bf k}}\sum_{\nu=1,2}(-1)^{\nu+1}{B^{(\nu)}_{{\rm z}{\bf k}}\over 2\omega^{(\nu)}_{{\rm z}{\bf k}}}\nonumber\\
&&\times\left ({1\over \omega-\omega^{(\nu)}_{{\rm z} {\bf k}}}- {1\over \omega+\omega^{(\nu)}_{{\rm z}{\bf k}}} \right ), \label{MF-spin-GF-inter-part-2}
\end{eqnarray}
\end{subequations}
with $B^{(\nu)}_{\bf k}=\lambda[(-1)^{\nu+1}A_{1}|\gamma_{\bf k}|-A_{2}]$, $B^{(\nu)}_{{\rm z}{\bf k}}=-\lambda\epsilon\chi[1-(-1)^{\nu+1}|\gamma_{\bf k}|]/2$, $\lambda=2ZJ_{\rm eff}$, $A_{1}=\epsilon\chi^{\rm z}+\chi/2$, $A_{2}=\chi^{\rm z}+\epsilon\chi/2$, the spin correlation function $\chi^{\rm z}=\langle S^{\rm z}_{{\rm A}l} S^{\rm z}_{{\rm B}l+\hat{\eta}} \rangle$,  $\epsilon=1+2t\phi/J_{\rm eff}$, the charge-carrier's particle-hole parameters $\phi=\langle h^{\dagger}_{{\rm A}l\sigma}h_{{\rm B}l+\hat{\eta}\sigma}\rangle=\langle h^{\dagger}_{{\rm B}l\sigma}h_{{\rm A}l+\hat{\eta}\sigma}\rangle$, and the MF spin excitation spectra,
\begin{subequations}
\begin{eqnarray}
(\omega^{(\nu)}_{\bf k})^{2}&=&\lambda^{2}[\epsilon\alpha A_{1}(|\gamma|^{2}_{\bf k}-{1\over Z})+{1\over 2}\epsilon^{2}A_{3}+A_{4}] \nonumber\\
&&-(-1)^{\nu+1}\epsilon\lambda^{2}[\alpha A_{2} (1-{1\over Z})+{A_{3}\over 2}+A_{4}]|\gamma_{\bf k}|,~~~~~~~\\
(\omega^{(\nu)}_{{\rm z}{\bf k}})^{2}&=&\epsilon\lambda^{2}[\alpha\chi(|\gamma|^{2}_{\bf k}-{1\over Z})+\epsilon A_{3}]\nonumber\\
&&-(-1)^{\nu+1}\epsilon\lambda^{2}[\alpha\chi(1-{1\over Z} ) +\epsilon A_{3}]|\gamma_{\bf k}|,
\end{eqnarray}
\end{subequations}
where $A_{3}=\alpha C+(1-\alpha)/(2Z)$, $A_{4}=\alpha C^{\rm z}+(1-\alpha)/(4Z)$, and the spin correlation functions $C=(1/Z^{2})\sum_{{\hat{\eta}}{\hat{\eta}'}}\langle S^{+}_{{\rm A} l+{\hat{\eta}}}S^{-}_{{\rm A}l+\hat{\eta}'}\rangle$, $C^{\rm z}=(1/Z^{2})\sum_{{\hat{\eta}}{\hat{\eta}'}}\langle S^{\rm z}_{{\rm A}l+\hat{\eta}} S^{\rm z}_{{\rm A}l+\hat{\eta}'} \rangle$. An important decoupling parameter $\alpha$, which can be regarded as the vertex correction, has been introduced in order to satisfy the sum rule $\langle S^{+}_{\upsilon l}S^{-}_{\upsilon l}\rangle=1/2$ in the case without an AFLRO \cite{Feng15}.

For unconventional superconductivity, electron Cooper pairs are not bound together by electron-phonon interaction. The inelastic neutron scattering and resonant inelastic X-ray scattering experiments revealed that spin excitations are over a large part of momentum space with high intensity and exist across the entire doping range of the SC dome \cite{Fujita12,Dean15}, thereby establishing a clear connection between the pairing mechanism and spin excitations. Considering the spin excitations mediating charge-carrier paring mechanism, the kinetic-energy-driven SC mechanism \cite{Feng03,Feng15} has been proposed by Feng et al to investigate the SC properties of cuprate superconductors. It was demonstrated that the interaction between charge-carriers and spins directly from the kinetic energy in the $t$–$J$ model is quite strong, which can induce the SC-state in the particle–particle channel by the exchange of spin excitations in the higher power of the doping concentration \cite{Feng03,Feng15}. It has been claimed that AF spin fluctuations are also present in doped graphene \cite{Peres04,Paiva05}, which could potentially lead to the emergence of unconventional superconductivity. Here we extend the theory of kinetic-energy-driven superconductivity for cuprate superconductors to investigate superconductivity on the honeycomb lattice.

Following the Eliashberg’s strong-coupling formalism \cite{Eliashberg60,Scalapino66,Mahan81}, the full charge-carrier normal and anomalous Green's functions satisfy the following self-consistent equations,
\begin{subequations}\label{full-charge-Green-function-sce}
\begin{eqnarray}
g({\bf k},\omega)&=&g^{(0)}({\bf k},\omega)+g^{(0)}({\bf k},\omega)[\Sigma_{1}^{({\rm h})}({\bf k},\omega)g({\bf k},\omega)\nonumber\\
&&-\Sigma_{2}^{({\rm h})}({\bf k},\omega)\Gamma^{\dagger}({\bf k},\omega)],\label{full-charge-normal-Green-function-sce}\\
\Gamma^{\dagger}({\bf k},\omega)&=&g^{(0)}({\bf k},-\omega)[\Sigma_{1}^{({\rm h})}({\bf k},-\omega)\Gamma^{\dagger}({\bf k},\omega)\nonumber\\
&&+\Sigma_{2}^{({\rm h})\dagger}({\bf k},\omega)g({\bf k},\omega)],\label{full-charge-anomalous-Green-function-sce}
\end{eqnarray}
\end{subequations}
where the Green's functions and self energies are defined as matrices,
\begin{subequations}\label{charge-Green-functions-matrix}
\begin{eqnarray}
g({\bf k},\omega)&=&\left(
\begin{array}{cc}
g_{\rm intra}({\bf k},\omega) & g_{\rm inter}({\bf k},\omega) \\
g^{*}_{\rm inter}({\bf k},\omega) & g_{\rm intra}({\bf k},\omega)
\end{array} \right) \,,\label{normal-Green-function-matrix}\\
\Gamma^{\dagger}({\bf k},\omega)&=&\left(
\begin{array}{cc}
0 & \Gamma^{\dagger}_{\rm inter}({\bf k},\omega) \\
\Gamma^{\dagger*}_{\rm inter}({\bf k},\omega) & 0
\end{array} \right) \,,\label{anomalous-Green-function-matrix}
\end{eqnarray}
\end{subequations}
and
\begin{subequations}\label{self-energy-matrix}
\begin{eqnarray}
\Sigma_{1}^{({\rm h})}({\bf k},\omega)&=&\left(
\begin{array}{cc}
\Sigma^{({\rm h})}_{1,\rm intra}({\bf k},\omega) & \Sigma^{({\rm h})}_{1,\rm inter}({\bf k},\omega) \\
\Sigma^{({\rm h})*}_{1,\rm inter}({\bf k},\omega) & \Sigma^{({\rm h})}_{1,\rm intra}({\bf k},\omega)
\end{array} \right) \,,\label{self-nergy-ph-matrix}\\
\Sigma_{2}^{({\rm h})}({\bf k},\omega)&=&\left(
\begin{array}{cc}
0 & \Sigma^{({\rm h})}_{2,\rm inter}({\bf k},\omega) \\
\Sigma^{({\rm h})*}_{2,\rm inter}({\bf k},\omega) & 0
\end{array} \right) \, .\label{self-nergy-pp-matrix}
\end{eqnarray}
\end{subequations}
Here we focus solely on the NN pairing of the charge-carriers, thus the intra-sublattice part of the charge-carrier anomalous Green's function is set to zero. We have to point out that Eq. (\ref{full-charge-anomalous-Green-function-sce}) is not strictly satisfied as the diagonal terms of the right side is not equal to zero. Nonetheless, we express it in this form for the sake of convenience in calculating the inter-sublattice part of the charge-carrier anomalous Green's function (the off-diagonal terms).

The intra-sublattice and inter-sublattice parts of the self energies in the particle-hole channel are,
\begin{subequations}\label{self-energy-ph}
\begin{eqnarray}
\Sigma^{({\rm h})}_{1,\rm intra}({\bf k},i\omega_{n})&=&{1\over N^{2}}\sum_{\bf pq}{|\Lambda_{\bf q}}|^{2}{1\over\beta}\sum_{ip_{m}}g_{\rm intra}({\bf p}, ip_{m})\nonumber\\
&&\times\Pi_{\rm intra}({\bf p},{\bf q},{\bf k},ip_{m},i\omega_{n}), ~~~~\label{intra-part-self-energy-ph}\\
\Sigma^{({\rm h})}_{1,\rm inter}({\bf k},i\omega_{n})&=&{1\over N^{2}}\sum_{\bf pq}{(\Lambda_{\bf q}})^{2}{1\over\beta}\sum_{ip_{m}}g^{*}_{\rm inter}({\bf p}, ip_{m})\nonumber\\
&&\times\Pi^{*}_{\rm inter}({\bf p},{\bf q},{\bf k},ip_{m},i\omega_{n}), ~~~~~~~\label{inter-part-self-energy-ph}
\end{eqnarray}
\end{subequations}
where $N$ is the number of lattice sites and $\beta=k_{\rm B}T$, and the inter-sublattice self energy in the particle-particle channel is,
\begin{subequations}\label{self-energy-pp}
\begin{eqnarray}
\Sigma^{({\rm h})}_{2,\rm inter}({\bf k},i\omega_{n})&=&{1\over N^{2}}\sum_{\bf pq}{(\Lambda_{\bf q}})^{2}{1\over\beta}\sum_{ip_{m}}\Gamma_{\rm inter}({\bf p}, ip_{m})\nonumber\\
&&\times\Pi^{*}_{\rm inter}({\bf p},{\bf q},{\bf k},ip_{m},i\omega_{n}), ~~~~~~~\label{inter-part-self-energy-pp}
\end{eqnarray}
\end{subequations}
where $\Lambda_{\bf k}=Zt\gamma_{\bf k}$, and the corresponding intra-sublattice and inter-sublattice parts of the spin bubbles,
\begin{subequations}\label{spin-bubble}
\begin{eqnarray}
\Pi_{\rm intra}({\bf p},{\bf q},{\bf k},ip_{m},i\omega_{n})&=&{1\over\beta}\sum_{iq_{m}}D^{(0)}_{\rm intra}({\bf p}+{\bf q},ip_{m}+iq_{m})\nonumber\\
&\times& D^{(0)}_{\rm intra}({\bf k}+{\bf q},i\omega_{n}+iq_{m}),~~~~~~~~\\
\Pi_{\rm inter}({\bf p},{\bf q},{\bf k},ip_{m},i\omega_{n})&=&{1\over\beta}\sum_{iq_{m}}D^{(0)}_{\rm inter}({\bf p}+{\bf q},ip_{m}+iq_{m})\nonumber\\
&\times& D^{(0)}_{\rm inter}({\bf k}+{\bf q},i\omega_{n}+iq_{m}).
\end{eqnarray}
\end{subequations}

The self energy $\Sigma_{1}^{({\rm h})}({\bf k},\omega)$ in the particle-hole channel is not an even function of $\omega$, but can be separated into the symmetric and antisymmetric parts, i.e., $\Sigma_{1}^{({\rm h})}({\bf k},\omega)=\Sigma^{({\rm h})}_{1\rm e}({\bf k},\omega)+\omega\Sigma^{({\rm h})}_{1\rm o}({\bf k},\omega)$, with $\Sigma^{({\rm h})}_{1\rm e}({\bf k},\omega)$ and $\Sigma^{({\rm h})}_{1\rm o}({\bf k},\omega)$ both being even function of $\omega$. As shown for the case of cuprate superconductors \cite{Feng15}, the antisymmetric part is directly associated with the charge-carrier quasiparticle coherent weight. Following the discussion of cuprate superconductors, we define the charge-carrier quasiparticle coherent weights as,
\begin{eqnarray}\label{quasiparticle-coherent-weight}
Z^{(\nu)-1}_{\rm hF}({\bf k},\omega)=1-{\rm Re}\Sigma^{({\rm h})}_{1,{\rm o}\nu}({\bf k},\omega),
\end{eqnarray}
where $\nu=1$ and $\nu=2$ mark the bonding and anti-bonding cases, respectively, and the bonding and anti-bonding self energies in the particle-hole channel are defined as,
\begin{subequations}\label{bonding-antibonding-self-energy-ph}
\begin{eqnarray}
\Sigma^{({\rm h})}_{1,{\rm o}1}({\bf k},\omega)&=&\Sigma^{({\rm h})}_{1,\rm intra}({\bf k},\omega)+e^{-i\theta_{\bf k}}\Sigma^{({\rm h})}_{1,\rm inter}({\bf k},\omega),\\
\Sigma^{({\rm h})}_{1,{\rm o}2}({\bf k},\omega)&=&\Sigma^{({\rm h})}_{1,\rm intra}({\bf k},\omega)-e^{-i\theta_{\bf k}}\Sigma^{({\rm h})}_{1,\rm inter}({\bf k},\omega),~~~~~
\end{eqnarray}
\end{subequations}
respectively. The self energy $\Sigma^{({\rm h})}_{2}({\bf k}, \omega)$ in the particle-particle channel incorporates the charge-carrier pairing force and pair order parameter, leading to the definition of the charge-carrier pair gap as,
\begin{eqnarray}
{\bar \Delta}_{{\rm h}}({\bf k},\omega)=e^{-i\theta_{\bf k}}\Sigma^{({\rm h})}_{2,{\rm inter}}({\bf k},\omega),
\end{eqnarray}
which represents the energy for breaking a charge-carrier pair. In the case of studying kinetic-energy-driven superconductivity we can adopt the static limit, i.e., ${\bar \Delta}_{\rm h}({\bf k})=e^{-i\theta_{\bf k}}\Sigma^{({\rm h})}_{2,{\rm inter}}({\bf k},\omega=0)$ and $Z^{(\nu)-1}_{\rm hF}({\bf k})=1-{\rm Re}\Sigma^{({\rm h})}_{1,{\rm o}\nu}({\bf k},\omega=0)$. As previously mentioned, numerous studies \cite{Honerkamp08,Raghu10,Kiesel12,Wu13,Song-Jin21,Li16,Pathak10,Wang12,Jiang14,Yu22,Ma11,Black-Schaffer14,Ying18,Guo19,Ying20,Ying22,Jiang14,Black-Schaffer14,Gu13,Jiang08,Nandkishore12nphy,Xu16,Plakida19,Khee-Kyun20,Ribeiro23,Khee-Kyun23} claimed that the $d+{\rm i}d$-wave pairing symmetry dominates the SC state on the honeycomb lattice. Therefore, we will focus on the $d+{\rm i}d$-wave SC gap here. In this case, the charge-carrier pair gap can be written as ${\bar \Delta}_{{\rm h}}({\bf k})={\bar \Delta}_{{\rm h}}d_{\bf k}={\bar \Delta}_{{\rm h}}(d_{1\bf k}+{\rm i}d_{2\bf k})$, with $d_{1\bf k}=\{\cos(k_{x}/2)[\sqrt{3}\sin(\sqrt{3}k_{y}/2)-\cos(\sqrt{3}k_{y}/2)]+\cos{k_{x}}\}/3$ and $d_{2\bf k}=\{\sin(k_{x}/2)[\sqrt{3}\sin(\sqrt{3}k_{y}/2)-\cos(\sqrt{3}k_{y}/2)]-\sin{k_{x}}\}/3$. The specific momentum dependence of $Z_{\rm hF}^{(\nu)}({\bf k})$ may not be crucial for qualitative analysis, so the wave vector ${\bf k}$ can be chosen as ${\bf k}_{0}=[2\pi/3,0]$ to evaluate $Z_{\rm hF}^{(\nu)}$, i.e.,
\begin{eqnarray}\label{quasiparticle-coherent-weight-k0}
Z^{(\nu)-1}_{\rm hF}=1-{\rm Re}\Sigma^{({\rm h})}_{1,{\rm o}\nu}({\bf k},\omega=0)\mid_{{\bf k}_{0}}.
\end{eqnarray}
After conducting the above process, the full charge-carrier normal and anomalous Green's functions are obtained as,
\begin{subequations}\label{full-charge-Green-function}
\begin{eqnarray}
g_{\rm intra}({\bf k},\omega)&=&{1\over 2}\sum_{\nu=1,2}Z^{(\nu)}_{\rm hF}\left [{(U_{{\rm h}{\bf k}}^{(\nu)})]^{2}\over \omega -E_{{\rm h}{\bf k}}^{(\nu)}}+{(V_{{\rm h}{\bf k}}^{(\nu)})^{2}\over \omega+E_{{\rm h}{\bf k}}^{(\nu)}}\right ], ~~~~~~~~\label{full-intra-charge-normal-Green-function}\\
g_{\rm inter}({\bf k},\omega)&=&{1\over 2}e^{i\theta_{\bf k}}\sum_{\nu=1,2}(-1)^{\nu+1}Z^{(\nu)}_{\rm hF}\nonumber\\
&&\times\left [{(U_{{\rm h}{\bf k}}^{(\nu)})^{2}\over \omega-E_{{\rm h}{\bf k}}^{(\nu)}}+{(V_{{\rm h}{\bf k}}^{(\nu)})^{2}\over \omega+E_{{\rm h}{\bf k}}^{(\nu)}}\right ],\label{full-inter-charge-normal-Green-function}\\
\Gamma^{\dagger}_{\rm inter}({\bf k},\omega)&=-&{1\over 2}e^{i\theta_{\bf k}}\sum_{\nu=1,2}(-1)^{\nu+1}Z^{(\nu)}_{\rm hF}{{\bar\Delta}_{\rm hZ}^{(\nu)*}({\bf k})\over 2E_{{\rm h}{\bf k}}^{(\nu)}}\nonumber\\
&&\times\left [{1\over \omega-E_{{\rm h}{\bf k}}^{(\nu)}}-{1\over \omega+E_{{\rm h}{\bf k}}^{(\nu)}}\right ],\label{full-inter-charge-anomalous-Green-function}
\end{eqnarray}
\end{subequations}
where $E_{{\rm h}{\bf k}}^{(\nu)}=\sqrt{[{\bar\xi}^{(\nu)}_{\bf k}]^{2}+|{\bar\Delta}_{\rm hZ}^{(\nu)}({\bf k})|^{2}}$ are charge-carrier quasiparticle energy spectra, with $\bar{\xi}^{(\nu)}_{\bf k}=Z^{(\nu)}_{\rm hF}\xi^{(\nu)}_{\bf k}$ and ${\bar\Delta}_{\rm hZ}^{(\nu)}({\bf k})=Z^{(\nu)}_{\rm hF}{\bar\Delta}_{\rm h}({\bf k})$, and $(U_{{\rm h}{\bf k}}^{(\nu)})^{2}$ and $(V_{{\rm h}{\bf k}}^{(\nu)})^{2}$ are charge-carrier quasiparticle coherence factors,
\begin{subequations}\label{charge-coherence-factors}
\begin{eqnarray}
{(U_{{\rm h}{\bf k}}^{(\nu)})^{2}}&=&{1\over 2}\left(1+{\bar{\xi}^{(\nu)}_{\bf k}\over E_{{\rm h}{\bf k}}^{(\nu)}}\right),\label{charge-coherence-factor-u}\\
{(V_{{\rm h}{\bf k}}^{(\nu)})^{2}}&=&{1\over 2}\left(1-{\bar{\xi}^{(\nu)}_{\bf k}\over E_{{\rm h}{\bf k}}^{(\nu)}}\right),\label{charge-coherence-factor-v}
\end{eqnarray}
\end{subequations}
which naturally satisfy the normalization constrain $(U_{{\rm h}{\bf k}}^{(\nu)})^{2}+(V_{{\rm h}{\bf k}}^{(\nu)})^{2}=1$. With the help of full charge-carrier Green's functions (\ref{full-charge-Green-function}) and the MF spin Green's functions in Eqs. (\ref{MF-spin-GF-intra-part-1}) and (\ref{MF-spin-GF-inter-part-1}), the self energies in the particle-hole channel and particle-particle channel are obtained explicitly as,
\begin{widetext}
\begin{subequations}\label{specific-self-energies}
\begin{eqnarray}
\Sigma^{({\rm h})}_{1,\rm intra}({\bf k},\omega)&=&{1\over N^{2}}\sum_{\bf pq}\sum_{\nu_{1}\nu_{2}\nu_{3}}\sum_{\mu}(-1)^{\mu+1}\Omega^{(\nu_{1}\nu_{2}\nu_{3})}_{{\rm h}\bf pqk}\nonumber\\
&&\times\left \{[U_{{\rm h}{\bf p}}^{(\nu_{3})}]^{2}\left [{F^{(\nu_{1}\nu_{2}\nu_{3})}_{{\rm 1h} {\mu\bf pqk}}\over\omega+\omega_{\mu\bf pqk}^{(\nu_{1}\nu_{2})}-E_{{\rm h}{\bf p}}^{(\nu_{3})}}+ {F^{(\nu_{1}\nu_{2}\nu_{3})}_{{\rm 2h} {\mu\bf pqk}}\over\omega-\omega_{\mu\bf pqk}^{(\nu_{1}\nu_{2})} -E_{{\rm h}{\bf p}}^{(\nu_{3})}} \right ]\right . \nonumber\\
&&+\left . [V_{{\rm h}{\bf p}}^{(\nu_{3})}]^{2}\left [{F^{(\nu_{1}\nu_{2}\nu_{3})}_{{\rm 1h} {\mu\bf pqk}}\over\omega-\omega_{\mu\bf pqk}^{(\nu_{1}\nu_{2})}+E_{{\rm h}{\bf p}}^{(\nu_{3})}}+{F^{(\nu_{1}\nu_{2}\nu_{3})}_{{\rm 2h} {\mu\bf pqk}}\over\omega+\omega_{\mu\bf pqk}^{(\nu_{1}\nu_{2})} +E_{{\rm h}{\bf p}}^{(\nu_{3})}} \right ]\right \},\label{specific-intra-self-energ-ph}\\
\Sigma^{({\rm h})}_{1,\rm inter}({\bf k},\omega)&=&{1\over N^{2}}\sum_{\bf pq}\sum_{\nu_{1}\nu_{2}\nu_{3}}\sum_{\mu}(-1)^{\mu+\nu_{1}+\nu_{2}+\nu_{3}}e^{-{\rm i}(2\theta_{\bf q}+\theta_{\bf p}-\theta_{\bf k+q}-\theta_{\bf p+q} )}\Omega^{(\nu_{1}\nu_{2}\nu_{3})}_{{\rm h}\bf pqk}\nonumber\\
&&\times\left \{[U_{{\rm h}{\bf p}}^{(\nu_{3})}]^{2}\left [{F^{(\nu_{1}\nu_{2}\nu_{3})}_{{\rm 1h} {\mu\bf pqk}}\over\omega+\omega_{\mu\bf pqk}^{(\nu_{1}\nu_{2})}-E_{{\rm h}{\bf p}}^{(\nu_{3})}}+ {F^{(\nu_{1}\nu_{2}\nu_{3})}_{{\rm 2h} {\mu\bf pqk}}\over\omega-\omega_{\mu\bf pqk}^{(\nu_{1}\nu_{2})} -E_{{\rm h}{\bf p}}^{(\nu_{3})}} \right ]\right . \nonumber\\
&&+\left . [V_{{\rm h}{\bf p}}^{(\nu_{3})}]^{2}\left [{F^{(\nu_{1}\nu_{2}\nu_{3})}_{{\rm 1h} {\mu\bf pqk}}\over\omega-\omega_{\mu\bf pqk}^{(\nu_{1}\nu_{2})}+E_{{\rm h}{\bf p}}^{(\nu_{3})}}+{F^{(\nu_{1}\nu_{2}\nu_{3})}_{{\rm 2h} {\mu\bf pqk}}\over\omega+\omega_{\mu\bf pqk}^{(\nu_{1}\nu_{2})} +E_{{\rm h}{\bf p}}^{(\nu_{3})}} \right ]\right \},\label{specific-inter-self-energ-ph}\\
\Sigma^{({\rm h})}_{2,\rm inter}({\bf k},\omega)&=&{1\over N^{2}}\sum_{\bf pq}\sum_{\nu_{1}\nu_{2}\nu_{3}}\sum_{\mu}(-1)^{\mu+\nu_{1}+\nu_{2}+\nu_{3}+1}e^{{\rm i}(2\theta_{\bf q}+\theta_{\bf p}-\theta_{\bf k+q}-\theta_{\bf p+q} )}\Omega^{(\nu_{1}\nu_{2}\nu_{3})}_{{\rm h}\bf pqk}{\bar{\Delta}_{\rm hZ}^{(\nu)}({\bf p})\over 2E_{{\rm h}{\bf p}}^{(\nu_{3})}}\nonumber\\
&&\times\left \{\left [{F^{(\nu_{1}\nu_{2}\nu_{3})}_{{\rm 1h} {\mu\bf pqk}}\over\omega+\omega_{\mu\bf pqk}^{(\nu_{1}\nu_{2})}-E_{{\rm h}{\bf p}}^{(\nu_{3})}}\right. +{F^{(\nu_{1}\nu_{2}\nu_{3})}_{{\rm 2h} {\mu\bf pqk}}\over\omega-\omega_{\mu\bf pqk}^{(\nu_{1}\nu_{2})} -E_{{\rm h}{\bf p}}^{(\nu_{3})}} \right ]\nonumber\\
&&-\left. \left [{F^{(\nu_{1}\nu_{2}\nu_{3})}_{{\rm 1h} {\mu\bf pqk}}\over\omega-\omega_{\mu\bf pqk}^{(\nu_{1}\nu_{2})}+E_{{\rm h}{\bf p}}^{(\nu_{3})}}+{F^{(\nu_{1}\nu_{2}\nu_{3})}_{{\rm 2h} {\mu\bf pqk}}\over\omega+\omega_{\mu\bf pqk}^{(\nu_{1}\nu_{2})} +E_{{\rm h}{\bf p}}^{(\nu_{3})}} \right ]\right \}, \label{specific-inter-self-energ-pp}
\end{eqnarray}
\end{subequations}
respectively, where $\mu=1,2$, $\Omega^{(\nu_{1}\nu_{2}\nu_{3})}_{{\rm h}\bf pqk}=Z_{\rm hF}^{(\nu_{3})}|\Lambda_{\bf q}|^{2}B_{\bf k+q}^{(\nu_{1})}B_{\bf p+q}^{(\nu_{2})}/[8\omega_{\bf k+q}^{(\nu_{1})}\omega_{\bf p+q}^{(\nu_{2})}]$, $\omega_{\mu\bf pqk}^{(\nu_{1}\nu_{2})}=\omega_{\bf k+q}^{(\nu_{1})}-(-1)^{\mu}\omega_{\bf p+q}^{(\nu_{2})}$, and
\begin{subequations}
\begin{eqnarray}
F^{(\nu_{1}\nu_{2}\nu_{3})}_{{\rm 1h} {\mu\bf pqk}}&=&n_{\rm F}(E_{{\rm h}{\bf p}}^{(\nu_{3})})\{1+n_{\rm B}(\omega_{\bf k+q}^{(\nu_{1})})+n_{\rm B}[(-1)^{\mu+1}\omega_{\bf p+q}^{(\nu_{2})}]\}+ n_{\rm B}(\omega_{\bf k+q}^{(\nu_{1})}) n_{\rm B}[(-1)^{\mu+1}\omega_{\bf p+q}^{(\nu_{2})}], \\
F^{(\nu_{1}\nu_{2}\nu_{3})}_{{\rm 2h} {\mu\bf pqk}}&=&[1-n_{\rm F}(E_{{\rm h}{\bf p}}^{(\nu_{3})})]\{1+n_{\rm B}(\omega_{\bf k+q}^{(\nu_{1})})+n_{\rm B}[(-1)^{\mu+1}\omega_{\bf p+q}^{(\nu_{2})}]\}+ n_{\rm B}(\omega_{\bf k+q}^{(\nu_{1})}) n_{\rm B}[(-1)^{\mu+1}\omega_{\bf p+q}^{(\nu_{2})}],
\end{eqnarray}
\end{subequations}
where $n_{\rm F}(\omega)$ and $n_{\rm B}(\omega)$ are fermion and boson distribution functions, respectively. Thus, according to Eqs. (\ref{specific-self-energies}) the charge-carrier quasiparticle coherent weight $Z_{\rm hF}^{(\nu)}$ and charge-carrier pair gap parameter $\bar{\Delta}_{\rm h}$ satisfy the following self-consistent equations,
\begin{subequations}\label{self-consistent-equations-1}
\begin{eqnarray}
{1\over Z_{\rm hF}^{(\nu)}}&=&1+{1\over N^{2}}\sum_{\bf pq}\sum_{\nu_{1}\nu_{2}\nu_{3}}\sum_{\mu}(-1)^{\mu+1}[1+(-1)^{\nu+\nu_{1}+\nu_{2}+\nu_{3}}\cos(2\theta_{\bf q}+\theta_{\bf p}+\theta_{{\bf k}_{0}}-\theta_{{\bf k}_{0}+{\bf q}}-\theta_{\bf p+q} )]\nonumber\\
&&\times\Omega^{(\nu_{1}\nu_{2}\nu_{3})}_{{\rm h}{\bf pqk}_{0}}\left \{{F^{(\nu_{1}\nu_{2}\nu_{3})}_{{\rm 1h} {\mu{\bf pqk}_{0}}}\over [\omega_{\mu{\bf pqk}_{0}}^{(\nu_{1}\nu_{2})}-E_{{\rm h}{\bf p}}^{(\nu_{3})}]^{2}} + {F^{(\nu_{1}\nu_{2}\nu_{3})}_{{\rm 2h} {\mu{\bf pqk}_{0}}}\over [\omega_{\mu{\bf pqk}_{0}}^{(\nu_{1}\nu_{2})} +E_{{\rm h}{\bf p}}^{(\nu_{3})}]^{2}} \right \},\\
1&=&{6\over N^{3}}\sum_{\bf pqk}\sum_{\nu_{1}\nu_{2}\nu_{3}}\sum_{\mu}(-1)^{\mu+\nu_{1}+\nu_{2}+\nu_{3}+1}\cos(2\theta_{\bf q}+\theta_{\bf p}+\theta_{\bf k}-\theta_{{\bf k}+{\bf q}}-\theta_{{\bf p}+{\bf q}} )\nonumber\\
&&\times\Omega^{(\nu_{1}\nu_{2}\nu_{3})}_{{\rm h}\bf pqk}{d_{1\bf k}d_{1\bf p}\over E_{{\rm h}{\bf p}}^{(\nu_{3})}}\left \{{F^{(\nu_{1}\nu_{2}\nu_{3})}_{{\rm 1h} {\mu\bf pqk}}\over \omega_{\mu\bf pqk}^{(\nu_{1}\nu_{2})}-E_{{\rm h}{\bf p}}^{(\nu_{3})}} + {F^{(\nu_{1}\nu_{2}\nu_{3})}_{{\rm 2h} {\mu\bf pqk}}\over \omega_{\mu\bf pqk}^{(\nu_{1}\nu_{2})} +E_{{\rm h}{\bf p}}^{(\nu_{3})}} \right \},
\end{eqnarray}
\end{subequations}
\end{widetext}
Eqs. (\ref{self-consistent-equations-1}) must be solved with other self-consistent equations,
\begin{subequations}\label{self-consistent-equations-2}
\begin{eqnarray}
\delta &=& {1\over 4N}\sum_{\nu,{\bf k}}Z^{(\nu)}_{\rm hF}\left ( 1-{\bar{\xi}^{(\nu)}_{\bf k}\over E^{(\nu)}_{{\rm h}\bf k}}{\rm tanh}[{1\over 2}\beta E^{(\nu)}_{{\rm h}\bf k}]\right ),
\end{eqnarray}
\begin{eqnarray}
\phi &=& {1\over 4N}\sum_{\nu,{\bf k}}(-1)^{\nu+1}|\gamma_{\bf k}|Z^{(\nu)}_{\rm hF}\nonumber\\
&&\times\left ( 1-{\bar{\xi}^{(\nu)}_{\bf k}\over E^{(\nu)}_{{\rm h}\bf k}}{\rm tanh}[{1\over 2}\beta E^{(\nu)}_{{\rm h}\bf k}]\right ),~~~~~~~~\\
{1\over 2} &=& {1\over 2N}\sum_{\nu,{\bf k}}{B^{(\nu)}_{\bf k}\over\omega^{(\nu)}_{\bf k}}{\rm coth} [{1\over 2}\beta\omega^{(\nu)}_{\bf k}],\\
\chi &=& {1\over 2N}\sum_{\nu,{\bf k}}(-1)^{\nu+1}|\gamma_{\bf k}|{B^{(\nu)}_{\bf k}\over\omega^{(\nu)}_{\bf k}}{\rm coth} [{1\over 2}\beta\omega^{(\nu)}_{\bf k}],\\
C &=& {1\over 2N}\sum_{\nu,{\bf k}}|\gamma_{\bf k}|^{2}{B^{(\nu)}_{\bf k}\over\omega^{(\nu)}_{\bf k}}{\rm coth} [{1\over 2}\beta\omega^{(\nu)}_{\bf k}],\\
\chi^{\rm z} &=& {1\over 2N}\sum_{\nu,{\bf k}}(-1)^{\nu+1}|\gamma_{\bf k}|{B^{(\nu)}_{{\rm z}{\bf k}}\over\omega^{(\nu)}_{{\rm z}{\bf k}}}{\rm coth} [{1\over 2}\beta \omega^{(\nu)}_{{\rm z}{\bf k}}],\\
C^{\rm z}&=&{1\over 2N}\sum_{\nu,{\bf k}}|\gamma_{\bf k}|^{2}{B^{(\nu)}_{{\rm z}{\bf k}}\over\omega^{(\nu)}_{{\rm z}{\bf k}}}{\rm coth} [{1\over 2}\beta\omega^{(\nu)}_{{\rm z}{\bf k} }].
\end{eqnarray}
\end{subequations}
Then the decoupling parameter $\alpha$, chemical potential $\mu$ and all the order parameters can be calculated by the self-consistent calculations.

\begin{figure}
\includegraphics[scale=0.35]{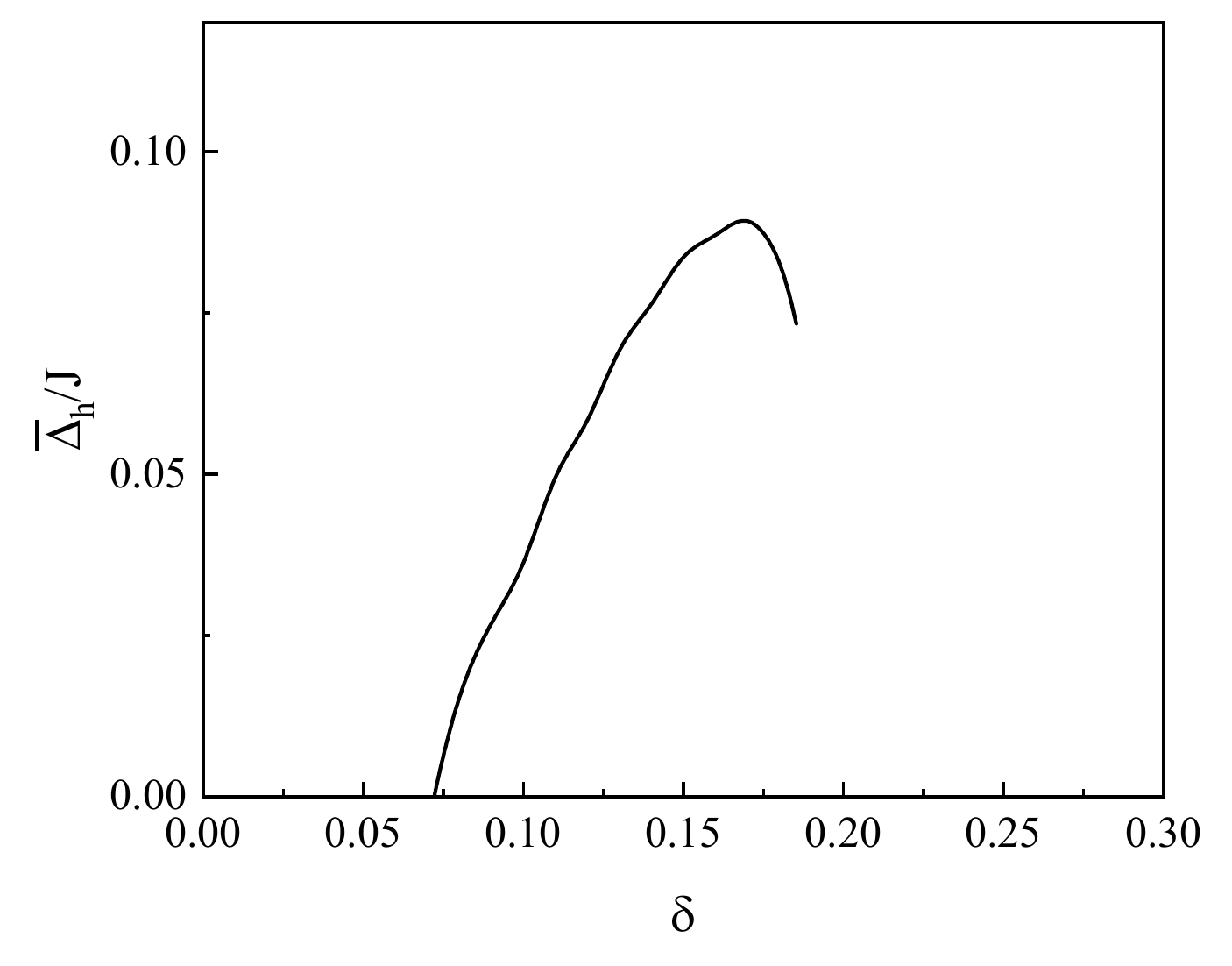}
\caption{Doping dependence of the charge-carrier pair gap parameter with $T=0.002J$ and
$t/J = 2.5$. \label{gap-parameter-vs-doping}}
\end{figure}

After the self-consistent calculations, the charge-carrier pair gap parameter ${\bar\Delta}_{\rm h}$ has been obtained and is plotted as a function of doping in Fig. \ref{gap-parameter-vs-doping} with $T=0.002J$ and $t/J = 2.5$. From Fig. \ref{gap-parameter-vs-doping} we can clearly see that, the doping dependence of charge-carrier pair gap parameter ${\bar\Delta}_{\rm h}$ exhibits a dome-like shape, being similar to the cases of cuprate superconductors \cite{Feng15} and cobaltate superconductors \cite{Feng18}. For small doping concentration lower than 0.07, ${\bar\Delta}_{\rm h}$ is equal to zero, implying the absence of superconductivity in this doping range probably due to the presence of strong AFLRO. As the doping concentration exceeds 0.072, ${\bar\Delta}_{\rm h}$ increases with increasing the doping concentration, reflecting the emergence of superconductivity. ${\bar\Delta}_{\rm h}$ reaches the maximal value around $\delta=0.17$ and then decreases with increasing the doping concentration, indicating that superconductivity will disappear at high doping levels. The dome-like shape of ${\bar\Delta}_{\rm h}$ with doping is partially consistent with other theoretical results \cite{Wu13,Gu13,Black-Schaffer14,Zhong15,Ribeiro23,Khee-Kyun23}, however, our results demonstrate the absence of superconductivity at low doping level, which is somewhat different with these theoretical studies \cite{Wu13,Gu13,Black-Schaffer14,Zhong15,Ribeiro23,Khee-Kyun23} that show the presence of superconductivity quite close to the region of half filling.

\begin{figure}
\includegraphics[scale=0.35]{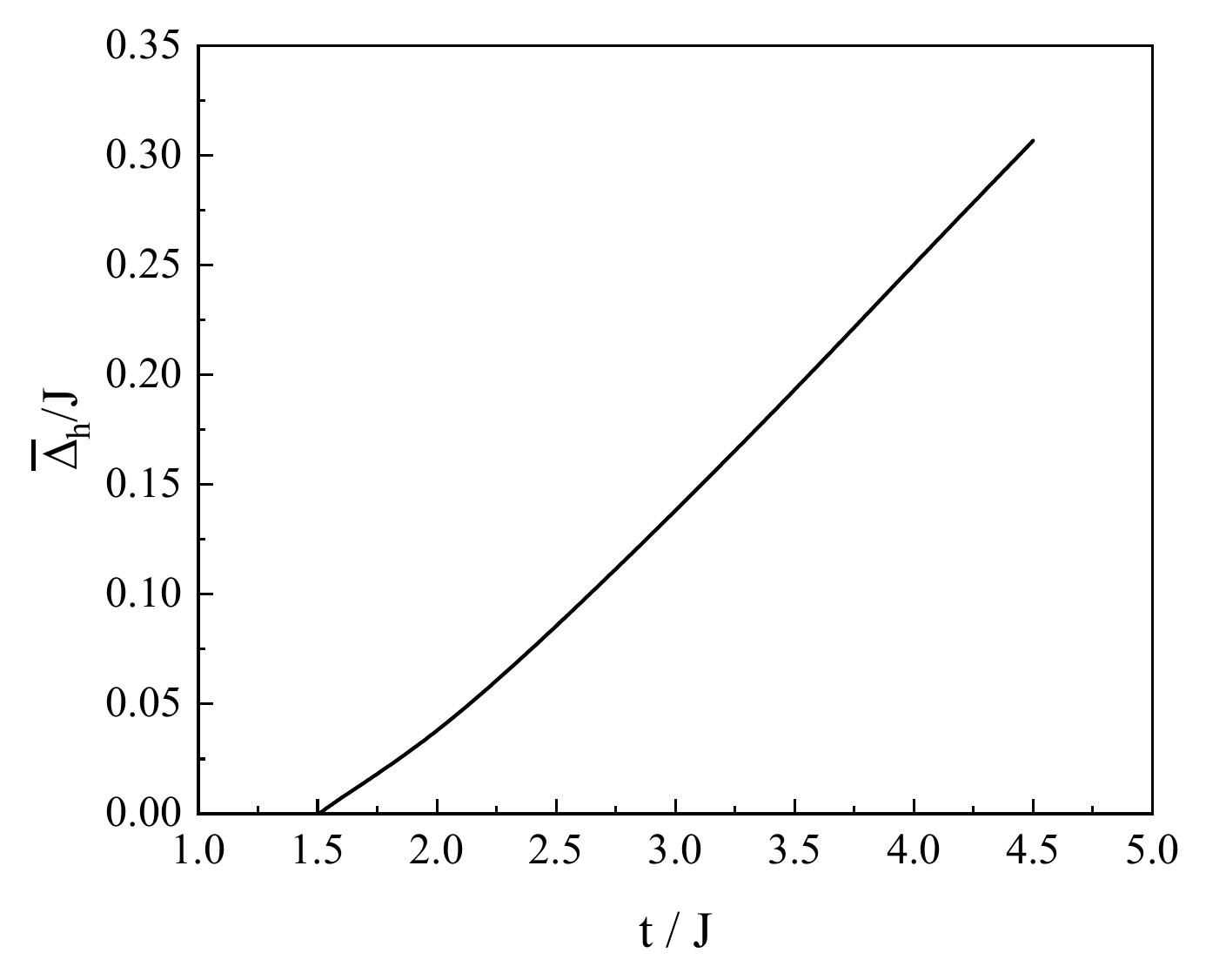}
\caption{The charge-carrier pair gap parameter as a function of $t/J$ with $T=0.002J$ at doping
$\delta = 0.15$. \label{gap-parameter-vs-tJ}}
\end{figure}

To see more clearly the superconductivity on a honeycomb lattice, we have investigated the $t/J$ dependence of the charge-carrier pair gap parameter and the result is plotted in Fig. \ref{gap-parameter-vs-tJ} with $T=0.002J$ at doping $\delta = 0.15$. Apparently, the charge-carrier pair gap parameter ${\bar\Delta}_{\rm h}$ increases monotonically with increasing the value of $t/J$. Large value of $t/J$ corresponds to small value of $J/t$, hence the charge-carrier pair gap parameter decreases with increasing the value of $J/t$. For small value of $J/t$, the spin-spin AF exchange coupling is suppressed, with ${\bar\Delta}_{\rm h}$ maintaining a rather large value, so the SC state is rather stable. As $J/t$ increases, the spin-spin AF order will be enhanced thus destroying the stability of the SC state. When $J/t \geq 2/3$ (corresponding to $t/J \leq 1.5$), the charge-carrier pair gap parameter ${\bar\Delta}_{\rm h}$ tends to zero, leading to the disappearing of superconductivity. On the other hand, $t/J \leq 1.5$ corresponds to $U/t = 4t/J \leq 6.0$, in a strength range which is probably beyond the condition of strong-coupling limit, leading to the invalidity of the $t$-$J$ model.

In conclusion, we have provided evidences supporting the possibility of unconventional superconductivity on a honeycomb lattice by employing the framework of kinetic-energy-driven SC mechanism. Within this scenario, spin excitations play a crucial role for the paring mechanism of the charge-carriers. The interaction between charge-carriers and spins directly from the kinetic energy in the $t$–$J$ model is rather strong, thus inducing the SC-state by the exchange of spin excitations in the higher power of the doping concentration. As a result, the $d+{\rm i}d$-wave charge-carrier pair gap parameter exhibits nonzero value within a certain doping range. However, in the low doping level superconductivity is suppressed by the strong antiferromagnetism, while in the high doping range superconductivity disappears due to the absence of spin excitations caused by the introducing a large number of charge-carriers into the materials. Moreover, as the value of $J/t$ increases, the charge-carrier pair gap parameter decreases and eventually reaches zero at approximately $J/t=2/3$, reflecting that the SC state will be gradually suppressed and finally be destroyed completely by increasing AF exchange coupling. The dome-like shape of the charge-carrier pair gap parameter with doping is rather similar to the results of cuprate and cobaltate superconductors, indicating that such behavior may be a common feature of doped Mott insulators. Our work offers valuable insights for future investigations of unconventional superconductivity on a honeycomb lattice.

\section*{Acknowledgements}

This work is supported by the Research Project of Education Department of Hunan Province under Grant No. 20K017, and the Key Laboratory of Optoelectronic Control and Detection Technology, University of Hunan Province.


\bibliographystyle{apsrev4-1}
\bibliography{honeycomb_mplb_lan_ref}

\end{document}